\documentclass[twocolumn]{revtex4}
\usepackage{epsfig}
\usepackage{graphicx}
\usepackage{amssymb}
\usepackage{amsmath}
\usepackage{amsfonts}
\usepackage{mathrsfs}

\newcommand{\R}{\mathbb{R}}
\newcommand{\C}{\mathbb{C}}

\newcommand{\fn}{\mathfrak{n}}

\newcommand{\fz}{\mathfrak{z}}

\newcommand{\fK}{\mathfrak{K}}

\newcommand{\cR}{\mathcal{R}}

\newcommand{\be}{\begin{equation}}
\newcommand{\ee}{\end{equation}}
\newcommand{\bea}{\begin{eqnarray}}
\newcommand{\eea}{\end{eqnarray}}
\newcommand{\nn}{\nonumber}

\newcommand{\ed}{\end{document}}

\newcommand{\bi}{\begin{itemize}}
\newcommand{\ei}{\end{itemize}}

\newcommand{\bce}{\begin{center}}
\newcommand{\ece}{\end{center}}

\newcommand{\RE}{\,{\rm Re}}
\newcommand{\IM}{\,{\rm Im}}

\newcommand{\nuemp}{\nu_{_{\!\varnothing}}}

\begin{document}

\title{Optical Spectral Singularities as Threshold Resonances}

\author{Ali Mostafazadeh}
\address{Department of Mathematics, Ko\c{c} University, 34450 Sar{\i}yer,
Istanbul, Turkey\\ amostafazadeh@ku.edu.tr}

\begin{abstract}

Spectral singularities are among generic mathematical features of
complex scattering potentials. Physically they correspond to
scattering states that behave like zero-width resonances. For a
simple optical system, we show that a spectral singularity appears
whenever the gain coefficient coincides with its threshold value and
other parameters of the system are selected properly. We explore a
concrete realization of spectral singularities for a typical
semiconductor gain medium and propose a method of constructing a
tunable laser that operates at threshold gain.

\hspace{6cm}{Pacs numbers: 03.65.Nk,  42.25.Bs, 42.60.Da, 24.30.Gd}



\end{abstract}

\maketitle

Consider an infinite planar slab gain medium that is aligned along
the $x$-axis and the electromagnetic wave given by $\vec
E(z,t)=E\:e^{i(\fK z-\omega t)}\hat e_x$, where $E$ is a constant, $\fK$ is the propagation constant, and
$\hat e_x$ stands for the unit vector pointing along the positive
$x$-axis. It is easy to show that while traveling through the gain
medium the wave is amplified by a factor of $e^{gL}$, where  $L$ is
the width of the gain medium and $g$ is the gain coefficient. The
latter is related to the imaginary part $\kappa$ of the the complex
refractive index of the medium,
    \be
    \mathfrak{n}=\eta+i\kappa,
    \label{n=}
    \ee
and the wavelength $\lambda:=2\pi c/\omega$ according to
\cite{silfvast}:
    \be
    g=-\frac{4\pi\kappa}{\lambda}.
    \label{gain-coeff}
    \ee
Often, one places the gain medium between two mirrors to produce a
(Fabry-Perot) resonator. This extends the length of the path of the
wave through the gain medium and yields a much larger amplification
of the wave for the resonance frequencies of the resonator. 

In \cite{prl-2009,pra-2009b}, we have outlined an alternative
amplification effect that does not involve mirrors. For the system
we consider, it requires adjusting $L$ and $g$ so that the system
supports a spectral singularity. This is a generic mathematical
feature of complex scattering potentials \cite{ss-math} that
obstructs the completeness of the eigenfunctions of the
corresponding non-Hermitian Hamiltonian operator. Physically, a
spectral singularity is the energy of a scattering state that
behaves exactly like a zero-width resonance
\cite{prl-2009,zafar-09,longhi,samsonov}. In this letter, we first
reveal the relationship between spectral singularities and the
well-known laser threshold condition \cite{silfvast}, and then
explore the possibility of tuning the wavelength of the spectral
singularity by adjusting the pump intensity. This turns the system
into a tunable laser that operates at the threshold gain.

It is easy to show that
    \be
    \vec E(z,t)=E\:e^{-i\omega t}\psi(z)\hat e_x,~~~
    \vec B(z,t)=-i \omega^{-1}E\:e^{-i\omega t}\psi'(z)\hat e_y,
    \nn
    \ee
is a solution of Maxwell's equations for the above system provided
that $\hat e_y$ stands for the unit vector along the positive
$y$-axis, $\psi$ is a continuously differentiable solution of the
time-independent Schr\"odinger equation:
    \be
    -\psi''(z)+v(z)\psi(z)=k^2\psi(z),
    \label{sch-eq}
    \ee
$k:=\omega/c=2\pi/\lambda$,   
and $v$ is the complex barrier potential:
    \be
    v(z):=\left\{\begin{array}{ccc}
    k^2\hat\fz &{\rm for}& |z|<\frac{L}{2},\\
    0&{\rm for}& |z|\geq\frac{L}{2},\end{array}\right.~~~~~
    \hat\fz:=1-\fn^2.
    \label{eqz-32}
    \ee

Solving the Schr\"odinger equation (\ref{sch-eq}) we find
    \be
    \psi(z)=\left\{\begin{array}{ccc}
    A_- e^{i k z}+B_- e^{-i k z} & {\rm for} & z\leq
    -\mbox{\small$L/2$}\\
    A_0 e^{i \fn k z}+B_0 e^{-i \fn k z}& {\rm for} & |z|<
    \mbox{\small$L/2$}\\
    A_+ e^{i k z}+B_+ e^{-i k z} & {\rm for} & z\geq
    \mbox{\small$L/2$},
    \end{array}\right.
    \label{eg-fu}
    \ee
where $k\in\R^+$, $A_-$ and $B_-$ are free complex coefficients and
$A_0, B_0, A_+$ and $B_+$ are complex coefficients related to $A_-$
and $B_-$. For example, $\mbox{\scriptsize$\left(\begin{array}{c}
A_+\\B_+\end{array}\right)$}=\mathbf{M}\mbox{\scriptsize$\left(\begin{array}{c}
A_-\\B_-\end{array}\right)$}$, where
    {\small\be
     \mathbf{M}:=\frac{1}{4\fn}\left(\begin{array}{cc}
    e^{-iL k}f(\fn,-\frac{Lk}{2}) & 2i(\fn^2-1)\sin(\fn L k)\\
    -2i(\fn^2-1)\sin(\fn Lk)& e^{iL k}f(\fn,\frac{Lk}{2})
    \end{array}\right)
    \nn
    \ee}%
is the transfer matrix, and for all $z_1,z_2\in\C$,
    \be
    f(z_1,z_2):=e^{-2iz_1z_2}(1+z_1)^2-e^{2iz_1z_2}(1-z_1)^2.
    \label{eq-f=}
    \ee
Because $v$ is an even function of $z$, the left and right
reflection and transmission amplitudes coincide. They are
respectively given by $R=-M_{21}/M_{22}=M_{12}/M_{22}$ and
$T=1/M_{22}$, \cite{prl-2009}.

The spectral singularities are the $k^2$ values for which
$M_{22}=0$, \cite{prl-2009}, i.e., the real $k$ values satisfying
    \be
    f(\fn,\mbox{$\frac{Lk}{2}$})=0.
    \label{f=0}
    \ee
Because $f$ is a complex-valued function,  Eq.~(\ref{f=0}) is
equivalent to a pair of coupled real transcendental equations for
three unknown real variables $\RE(\hat\fz)$, $\IM(\hat\fz)$ and
$\alpha k$. In Ref.~\cite{pra-2009b}, we outline a method of
decoupling these equations. Here we give a more direct solution that
reveals some previously unknown aspects of the problem.

First, we use (\ref{eq-f=}) to express  (\ref{f=0}) in the form
    \be
    e^{-2i\fn L k}-\left(\frac{1-\fn}{1+\fn}\right)^2=0.
    \label{eqz-11}
    \ee
Noting that the boundaries of the gain region have a reflectivity of
    \be
    \cR:=\left(\frac{\fn-1}{\fn+1}\right)^2,
    \label{reflex}
    \ee
we can write (\ref{eqz-11}) as $e^{-2iLk \fn }=\cR$. Next, we
substitute (\ref{n=}) in this equation and use (\ref{gain-coeff}) to
express its left-hand side in terms of $g$. This gives
    \be
    e^{-2gL}e^{-4iLk\eta}=\cR^2.
    \label{ss}
    \ee
Taking the modulus of both sides of this equation yields
    \be
     e^{-2gL}=|\cR|^2.
    \label{eqz-14}
    \ee
This is precisely the laser threshold condition \cite{silfvast}. In
other words, a spectral singularity and the associated zero-width
resonance appear at the threshold gain:
    \be
    g=g_{\rm th}:= \frac{1}{2L}\ln\frac{1}{|\cR|^2}.
    \label{eqz-15}
    \ee

We wish to stress that the threshold condition~(\ref{eqz-14}) is
only a necessary condition for having a spectral singularity. It is
by no means sufficient. A necessary and sufficient condition is
Eq.~(\ref{eqz-11}) that we can express as
    \be
    kL=-\frac{1}{2i\fn}\ln\cR.
    \label{gl=}
    \ee
A key observation that reveals the discrete nature of spectral
singularities is that, as a complex-valued function, $\ln\cR$ has
infinitely many values; in view of (\ref{n=}) and (\ref{reflex}),
    \be
    \ln\cR=\ln|\cR|+
    2i\left[\tan^{-1}\left(\frac{2\kappa}{\eta^2+\kappa^2-1}
    \right)-\pi m\right],~~~~
    \label{R=2}
    \ee
where $m$ is an arbitrary integer. Using (\ref{n=}) and
(\ref{reflex}), we also find
    \be
    |\cR|=\frac{(\eta-1)^2+\kappa^2}{
    (\eta+1)^2+\kappa^2}.
    \label{R=1}
    \ee
Because $\eta>0$, this equation implies $|\cR|<1$. Furthermore, in
view of (\ref{eqz-15}), we have
    \be
    g=g_{\rm th}= \frac{1}{L}\ln\left[\frac{(\eta+1)^2+\kappa^2}{
    (\eta-1)^2+\kappa^2}\right].
    \label{eqz-91}
    \ee

Next, we return to Eq.~(\ref{gl=}). Because $kL$ is real, the
imaginary part of the right-hand side of this equation must vanish.
This gives
    \be
    \eta\ln|\cR|+2\kappa\left[\tan^{-1}\left(\frac{2\kappa}{\eta^2+\kappa^2-1}
    \right)-\pi m\right]=0.
    \label{eqz71}
    \ee
Furthermore, we can express (\ref{gl=}) as
    \be
    kL=\frac{\ln|\cR|}{2\kappa}.
    \label{eqz72}
    \ee
Because $kL>0$ and $|\cR|<1$, (\ref{eqz72}) implies $\kappa<0$.
According to (\ref{gain-coeff}), this corresponds to the situation
that the medium has a positive gain coefficient. This is a
remarkable manifestation of the conservation of energy, because
whenever we arrange the parameters of the system so that a spectral
singularity is generated, the system begins emitting radiation. This
can happen only for a gain medium, i.e., when $g>0$. In
\cite{pra-2009b}, we could only demonstrate this graphically. Here
we have derived it rigorously.

Eq.~(\ref{eqz71}) determines the location of spectral singularities
in the $\eta$-$\kappa$ plane. In view of the inequalities: $\eta>0$,
$\kappa<0$, $|\cR|<1$, and the fact that $\tan^{-1}$ is an odd
function taking values in
$\left(-\frac{\pi}{2},\frac{\pi}{2}\right)$, we can satisfy
(\ref{eqz71}) only for $m\geq 0$. It is also instructive to note
that solving for $\ln|\cR|$ in (\ref{eqz71}), substituting the
result in (\ref{eqz72}), and using $k=2\pi\nu/c$, we find the
following expression for the frequency $\nu$ of the spectral
singularity.
    \bea
    \nu&=&\frac{\nuemp m}{\eta}-
    \frac{\nuemp}{\pi \eta} \tan^{-1}\left(
    \frac{2\kappa}{\eta^2+\kappa^2-1}\right),
    \label{frequency}
    \eea
where $\nuemp:=\frac{c}{2L}$. The first term on the right-hand side
of (\ref{frequency}) is the usual resonance frequency of a resonator
of length $L$ with perfectly reflecting boundaries. It is important
to note that because $\kappa$ and $\eta$ are frequency-dependent
quantities, there will be certain frequencies for which they satisfy
(\ref{eqz71}). A spectral singularity will arise if and only if at
least one of these frequencies coincides with one of the values
fulfilling (\ref{frequency}). It turns out that if we fix all the
physical parameters of the system this can happen only for a single
critical frequency, i.e., a particular mode number $m$.

\begin{figure}\vspace{-.5cm}
    \begin{center}
    \includegraphics[scale=.70,clip]{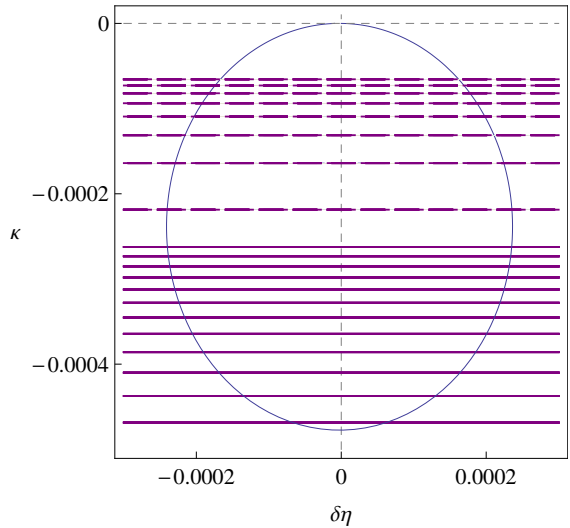}
    \vspace{-.3cm}
    \caption{(Color online)  Plots of the curves of spectral singularities $C_m$ (full and dashed thick purple  curves) and the dispersion relation $C$ (thin blue curve)  for a gain medium with specifications~(\ref{specific}). The spectral label $m$ of the $C_m$ that are displayed ranges over $1400, 1500, 1600, \cdots, 2500$ for the thick full (purple) curves from bottom to top, and $3000, 4000, 5000,\cdots,10000$ for the thick dashed (purple) curves. As one increases $m$ these curves approach to the $\delta\eta$-axis from below. The thin dashed (gray) lines are the coordinate axes. \label{fig2n}}
    \end{center}
    \end{figure}

In order to explore this phenomenon suppose that the gain medium is
obtained by doping a host medium of refraction index $n_0$ and that
it is modeled by a two-level atomic system with lower and upper
level population densities $N_l$ and $N_u$, resonance frequency
$\omega_0$, and damping coefficient $\gamma$. Then its permittivity
($\varepsilon:=\varepsilon_0\fn^2$) is given by
    \be
    \varepsilon= \varepsilon_0\left[n_0^2-
    \frac{\omega_p^2}{\omega^2-\omega_0^2+i\gamma\,\omega}\right],
    \label{epsilon}
    \ee
where $\varepsilon_0$ is the permittivity of the vacuum,
$\omega_p^2:=(N_l-N_u)e^2/(m\varepsilon_0)$, and $e$ and $m$ are
electron's charge and mass, respectively \cite{silfvast,yariv-yeh}.
In view of (\ref{n=}), (\ref{gain-coeff}), and (\ref{epsilon}),  we
can express $\omega_p^2$ in terms of the gain coefficient $g_0$ at
the resonance frequency $\omega_0$. Introducing $\lambda_0:=2\pi
c/\omega_0$ and
    \be
    \kappa_0:=-\frac{\lambda_0g_0}{4\pi},
    \label{k0}
    \ee
and using
    \be
    \varepsilon/\varepsilon_0=\fn^2=(\eta+i\kappa)^2,
     \label{eqz201}
     \ee
and (\ref{gain-coeff}), we have
   \be
   \frac{\omega_p^2}{\omega_0^2}=
   \frac{2\gamma\kappa_0\sqrt{n_0^2+\kappa_0^2}}{\omega_0}
   =-\frac{cg_0\gamma}{\omega_0^2}\sqrt{n_0^2+
   \left(\frac{cg_0}{2\omega_0}\right)^2}.
   \label{plasma}
    \ee
For example for a semiconductor gain medium \cite{silfvast} with:
   {\small \be
    n_0=3.4,~\lambda_0=1500\,{\rm nm},~
    \frac{\gamma}{\omega_0}=0.02,~g_0=40\,{\rm cm}^{-1},
    \label{specific}
    \ee}%
we have $\kappa_0=-4.7747\times 10^{-4}$ and
$\frac{\omega_p^2}{\omega_0^2}=-6.4935\times 10^{-5}$. Substituting
these values in (\ref{epsilon}) and using (\ref{eqz201}), we can
express $\delta\eta:=\eta-n_0$ and $\kappa$ as functions of $\omega$
and use them to plot a parametric curve $C$ representing the
dispersion relation~(\ref{epsilon}). See Figure~\ref{fig2n}. Next,
we return to Eq.~(\ref{eqz71}) and consider it as an implicit
function for each value of $m$. Figure~\ref{fig2n} also shows the
graph of these functions for several values of $m$. These are the
curves of spectral singularities that we label by $C_m$. As seen
from Figure~\ref{fig2n} for sufficiently large values of $m$ (i.e.,
$m\geq 1374$), $C_m$ intersects $C$ at two points. One of these
($p_{m+}$) has a positive $\delta\eta$ coordinate and corresponds to
a frequency $\omega_{m+}$ that is larger than $\omega_0$. The other
($p_{m+}$) has a negative $\delta\eta$ and corresponds to a
frequency $\omega_{m-}$ that is smaller than $\omega_0$. We can
compute the coordinates $(\delta\eta_{m\pm},\kappa_{m\pm})$ of
$p_{m\pm}$, find $\omega_{m\pm}$ and $k_{m\pm}:=\omega_{m\pm}/c$,
and use (\ref{eqz72}) to determine $L$. This gives rise to a set of
values $L_{m\pm}$ for the length of the gain medium that would allow
for the existence of  a spectral singularity. For the gain medium
(\ref{specific}), setting $m=1374$ we find a pair of spectral
singularities with extremely close wavelengths to $\lambda_0$,
namely $\lambda=1499.9$ and $1500.2\,{\rm nm}$. These correspond to
$L=303.074$ and $303.133\,\mu{\rm m}$, respectively.

Next, we consider the more realistic situation that $L$ is fixed and
the gain coefficient $g_0$ is adjustable. This is simply done by
changing the pump intensity. It is not difficult to see that
spectral singularities appear for specific discrete values of $g_0$
and $\lambda$. This may be viewed as means for producing a tunable
laser that would function at the very threshold gain. To implement
this idea we insert (\ref{plasma}) in (\ref{epsilon}) and use
(\ref{eqz201}) and $\lambda=2\pi c/\omega$ to express $\delta\eta$
and $\kappa$ as functions of $g_0$ and $\lambda$. Substituting the
resulting expressions in (\ref{eqz71}) and (\ref{eqz72}) and using
$k=2\pi/\lambda$ we obtain a pair of equations for $g_0$ and
$\lambda$ for each value of the mode number $m$. The solution of
these equations yield the desired values of $g_0$ and $\lambda$ for
which a spectral singularity appears.

As we see from Figure~\ref{fig2n}, the relevant range of values of
$\delta\eta$ and $\kappa$ is several orders of magnitude smaller
than $n_0$. This shows that we can obtain reliable approximate forms
of (\ref{eqz71}) and (\ref{eqz72}) by neglecting quadratic and
higher order terms in $\delta\eta$ and $\kappa$. Applying this
approximation to (\ref{eqz201}) and using (\ref{epsilon}), we have
    \be
    \delta\eta\approx-\kappa_0 f_1,~~~~
    \kappa\approx\kappa_0 f_2,
    \label{eqz301}
    \ee
where
    \be
    f_1:=\frac{1-\hat\omega^2}{(1-\hat\omega^2)^2+
    \hat\gamma^2\hat\omega^2},~~
    f_2:=\frac{\hat\gamma^2\hat\omega}{(1-\hat\omega^2)^2+
    \hat\gamma^2\hat\omega^2},
    \label{fs}
    \ee
$\hat\omega:=\omega/\omega_0=\lambda_0/\lambda$ and
$\hat\gamma:=\gamma/\omega_0$. In view of (\ref{eqz301}), the
above-mentioned approximation scheme is equivalent to neglecting
terms of order two and higher in $\kappa_0$.
Inserting~(\ref{eqz301}) in (\ref{eqz72}) and using
$k=2\pi/\lambda$, we find
    \be
    \kappa_0\approx-\frac{\rho_0}{2}\left(\frac{f_1}{n_0^2-1}+\frac{\pi L\hat\omega f_2}{\lambda_0}\right)^{-1},
    \label{eqz-401}
    \ee
where $\rho_0:=\ln[(n_0+1)/(n_0-1)]$. In light of (\ref{k0}),
(\ref{eqz-401}) is equivalent to
    \be
    g_0\approx 2\pi\rho_0\left(\frac{\lambda_0f_1}{n_0^2-1}+
    \pi L\hat\omega f_2\right)^{-1}.
    \label{eqz-402}
    \ee
Next, we substitute (\ref{eqz301}) in  (\ref{eqz72}), neglect the
quadratic and higher order terms in $\kappa_0$, and use
(\ref{eqz-401}) and (\ref{fs}) to simplify the result. This yields
after some lucky cancelations: $\left(\frac{2\pi L
n_0}{\lambda_0}-\frac{\rho_0}{\hat\gamma^2}\right) \hat\omega^2-\pi
m\,\hat\omega+\frac{\rho_0}{\hat\gamma^2}\approx 0$. Only for $m\geq
\mu:=\frac{2\rho_0}{\pi\hat\gamma}\sqrt{\frac{2\pi L
n_0}{\rho_0\lambda_0}-\frac{1}{\hat\gamma^2}}$ does this equation
have real solutions. These give the frequency (wavelength) of the
spectral singularities. Substituting them in (\ref{eqz-402}), we
find the corresponding $g_0$ values. The approximate values of
$\lambda$ and $g_0$ obtained in this way allow for a more effective
numerical solution of the exact equations (\ref{eqz71}) and
(\ref{eqz72}).

\begin{figure}
    \begin{center}
    \vspace{.5cm}
    \includegraphics[scale=.7,clip]{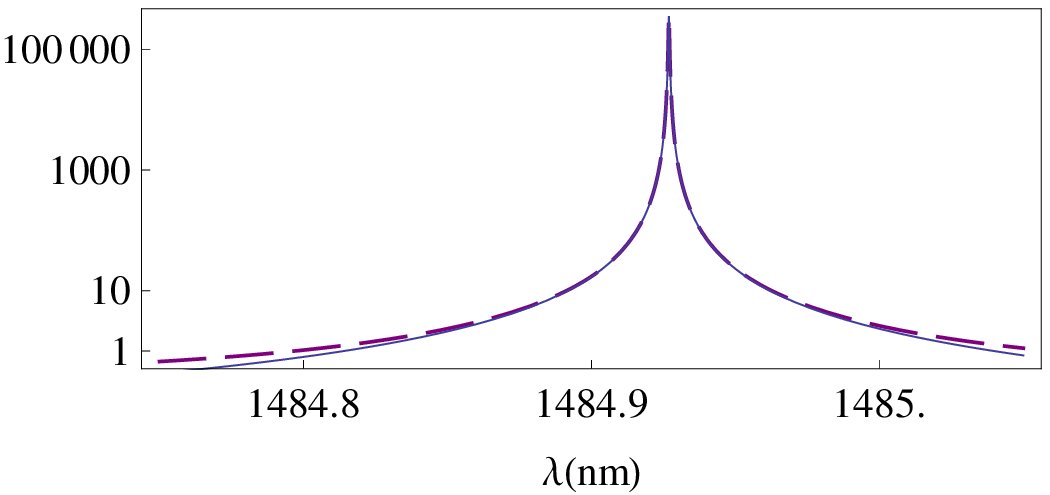}\\
    \includegraphics[scale=.7,clip]{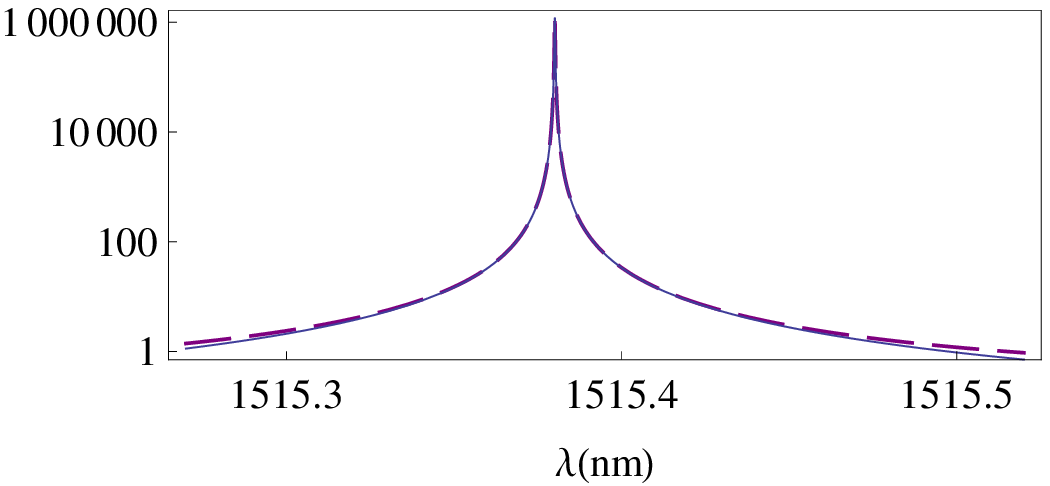}
    \vspace{-.3cm}
    \caption{(Color online)  Logarithmic plots of the reflection (thick dashed curves) and transmission (thin full curves) coefficients as functions of $\lambda$ for $L=300\,\mu{\rm m}$ and $g=81.7/{\rm cm}$ (top graph) and $g=82.5/{\rm cm}$ (bottom graph). \label{fig2}}
    \end{center}
    \end{figure}

As a concrete example, we take $L=300~\mu{\rm m}$ for the gain
medium described by (\ref{specific}) and obtain the wavelength of
the spectral singularities that are produced as we change $g_0$ in
the range $0$ - $115~{\rm cm}^{-1}$. These turn out to correspond to
the mode numbers $1355$ -- $1365$ which are consistent with the
result, $m\geq\mu\approx 1320$, of our approximate calculations.The
corresponding $\lambda$ and $g_0$ values are given in
Table~\ref{table1}. The  spectral singularity with lowest $g_0$
value ($40.4~{\rm cm}^{-1}$) has a wavelength that is extremely
close to $\lambda_0=1500\,{\rm nm}$. It corresponds to $m=1360$. As
we increase $g_0$ form $40.4$ to $115~{\rm cm}^{-1}$, there appear
10 more spectral singularities with wavelengths ranging between
$1482$ and $1520\,{\rm nm}$. These should in principle be
detectable, if we gradually increase the pump intensity. For example
a periodic change of $g_0$ in the range $70$ -- $90~{\rm cm}^{-1}$
should produce periodic emissions of radiation at the wavelengths
$\lambda=1484.927\,{\rm nm}$ (for $m=1364$ and $g_0=81.6821~{\rm
cm}^{-1}$) and $\lambda=1515.380\,{\rm nm}$ (for $m=1356$ and
$g_0=82.4918~{\rm cm}^{-1}$)  with no emitted wave of comparable
amplitude at the resonance wavelength $1500\,{\rm nm}$. This is a
remarkable feature of the spectral singularity related resonance
effect. Using the above values of $\lambda$ and $g_0$ to compute the
reflection and transmission coefficients $|R|^2$ and $|T|^2$, that
give the amplification factor for the emitted electromagnetic energy
density, we obtain $|R|^2\approx|T|^2\approx 1.1\times 10^6$ and
$3.1\times 10^5$, respectively. In other words we obtain an
amplification of the background electromagnetic energy density at
these wavelengths by a factor of $|R|^2+|T|^2\approx 2.2\times 10^6$
and $6.2\times 10^5$, respectively. It turns out that these numbers
are extremely sensitive to the value of $\lambda$ but not $g_0$.
Using the less accurate values $81.7$ and $82.5~{\rm cm}^{-1}$ for
$g_0$ and the same values for  $\lambda$, we find
$|R|^2\approx|T|^2\approx 1.0\times 10^6$ and $2.8\times 10^5$. But,
as we can see from Figure~\ref{fig2}, changing the above values of
$\lambda$ by $0.01$~{\rm nm} reduces $|R|^2$ and $|T|^2$ by three to
four orders of magnitude. This shows that the detection of spectral
singularities would require a spectrometer with a band width of
$0.01$~{\rm nm} or smaller.
    \begin{table}
    \begin{tabular}{|c|c|c||c|c|c|}\hline
  ~$m$~&$\lambda ({\rm nm})$ &$g_0 ({\rm cm}^{-1})$&
  ~$m$~&$\lambda ({\rm nm})$ &$g_0 ({\rm cm}^{-1})$\\
  \hline
 1355  & 1519.832 & 110.2 & 1361  & 1495.662 & 43.8 \\
 1356  & 1515.380 &  82.5  &  1362  & 1494.582 & 45.7\\
 1357  & 1512.058 &  66.3  &  1363  & 1489.203 & 61.5\\
 1358  & 1507.651 &  50.9  &  1364  & 1484.927 & 81.7\\
 1359  & 1504.363 & 43.8 &  1365  & 1481.736 & 101.1\\
 1360  & 1500.000 & 40.4 &  &&\\
 \hline
      \end{tabular}
  \caption{Wave length $\lambda$ of the spectral singularities for the gain coefficient $g_0$ ranging over $0$-$115~{\rm cm}^{-1}$ and $L=300~\mu{\rm m}$.
    \label{table1}}
\end{table}

To summarize, we have explored spectral singularities of a simple
optical system and shown that the equation (\ref{ss}) that
determines spectral singularities reduces to the laser threshold
condition $g=g_{\rm th}$, if we take the modulus of both sides of
this equation. Equating the phase of both sides of this equation
gives rise to an independent condition for the existence of the
spectral singularities that involves an integer (mode) number $m$.
It turns out that this equation and the threshold condition can be
satisfied only for particular values of the physical parameters of
the system and this corresponds to a single value of $m$ and a
corresponding critical wavelength. We have also explored the idea of
tuning this wavelength by adjusting the pump intensity for a typical
semiconductor gain medium. A remarkable feature of the spectral
singularity related resonance effect is that if we increase the pump
intensity so that the gain exceeds the threshold value, this effect
disappears. This marks a clear distinction between the zero-width
resonances associated with spectral singularities and the usual
resonances that we encounter in optical resonators.\vspace{.5cm}

\noindent {\em Acknowledgments:} I wish to thank Aref Mostafazadeh
and Ali Serpen\u{g}zel for illuminating discussions. This work has
been supported by the Turkish Academy of Sciences (T\"UBA).

\ed
\begin{thebibliography}{99}
\bibitem{silfvast} W.~T.~Silfvast, {\em Laser Fundamentals},
Cambridge University Press, Cambridge, 1996.

\bibitem{prl-2009} A.~Mostafazadeh, Phys.\ Rev.\ Lett.~\textbf{102}, 220402
(2009).

\bibitem{pra-2009b} A.~Mostafazadeh, Phys.\ Rev.\ A \textbf{80}, 032711
(2009).

\bibitem{ss-math} M.~A.~Naimark, Trudy Moscov.\ Mat.\ Obsc.\ \textbf{3}, 181 (1954) in Russian, Amer.\ Math.\ Soc.\ Transl.\
(2), \textbf{16}, 103 (1960); R.~R.~D.~Kemp, Canadian J. Math.
\textbf{10}, 447 (1958); J.~Schwartz, Comm.\ Pure Appl.\ Math.
\textbf{13}, 609 (1960); G.~Sh.~Guseinov, Pramana.\ J.~Phys.\
\textbf{73}, 587 (2009).
\bibitem{zafar-09} Z.~Ahmed, J.~Phys.~A \textbf{42}, 472005 (2009).
\bibitem{longhi} S.~Longhi, Phys.\ Rev.\ B  \textbf{80}, 165125 (2009) and
Phys.\ Rev.\ A \textbf{81}, 022102 (2010).
\bibitem{samsonov} B.~F.~Samsonov, preprint arXiv:1007.4421.
\bibitem{yariv-yeh} A.~Yariv and P.~Yeh, {\em Photonics,} Oxford
University Press, Oxford, 2007.



\end{thebibliography}
